\documentclass{jpsj2}

\def\runauthor{Online-Journal Subcommittee}

\title{%
Invisible Un-removable Field: A Search by Ultra-High Energy Cosmic Rays
}

\author{%
Humitaka Sato\thanks{E-mail address: satoh@konan-u.ac.jp}
}

\inst{%
Department of Physics, Konan University \\
Okamoto, Kobe 658-8501,JAPAN
}

\recdate{}

\abst{%
A Toy model which introduces the different limiting velocities for each 
particleis discussed: the differentiation of the universal limiting 
velocity is implemented throgh the coupling with external tensor or vector  
field, something like Higgs scalar field. GZK cut-off discussion could be 
altered due to the violation of Lorentz symmetry.  

}

\kword{%
relativity principle, High Energy cosmic rays, GZK-cutoff
}

\begin{document}
\maketitle

\section{Introduction and Summary}

One can characterize an achievement of the 20-century Physics as the 
discovery of various symmetries, those have been hidden deep in the 
diversity of superficial phenomena: we can point out many  symmetries 
such as rotational and boost  symmetry of 3-space, past-future symmetry 
in mechanics, duality symmetry between electro- and magneto-fields, 
Lorentz symmetry of spacetime, discrete symmetry in atomic structure of 
solid, particle-antiparticle symmetry,  isospin symmetry of nuclear force, 
chiral symmetry, "eight-fold symmetry", super-symmetry, colour symmetry 
and so on. Particularly, Standard Theory of elementary particles is 
formulated by the gauge-theory  based on internal or local symmetry 
hidden in electro-weak and strong interactions among quarks and leptons. 

However we have to remind that this unification is possible by introducing 
another extra idea called "spontaneous symmetry breakdown(SSB)". This 
mechanism is schematically written as
$${\rm [observed~~ law]=[symmetric~~ law]x[SSB]}.$$
New paradigm introduced here is that the observed law is separated into 
two different laws one is an ideal "symmetric law" and the another is to 
locate our universe in a@non-universal particular state. Therefore,  
the symmetric law itself can not make in existence in our universe.

In the Standard Theory, this SSB mechanism is controlled by the  external 
field called Higgs field: our universe has  been permeated by this external 
field and the interaction of particle with it gives a mass. Essential 
difference of the Higgs field from a conventional fields of particles is 
that it is un-removable from our universe. The Higgs external field is a 
"visible field" through the particle's mass.  

This SSB has introduced a new ingredient about the concept of  physics law, 
that is, the physics law itself is symmetric but our actual universe is 
not in a state of exact symmetry. This may be re-phrased also as followings; 
physics law is universal but our universe is not universal entity, or, 
physics law itself does not exhibit its original form in our universe 
where we live in. We call this kind of idea as the SSB paradigm.~\cite{b00} 
The SSB paradigm  describes our universe as "un-universal" universe.

In fact, some symmetries are not exact but show a tiny breakdown, like 
in case of CP-asymmetry. The actual composition of cosmic matter does not 
obey the particle-antiparticle symmetry in spite of CPT-symmetry in physics 
law itself. Following these considerations, we are tempted to think that 
any symmetry might be not exact in this actual universe, which has come 
into an existence through various spontaneous selections of non-universal 
parameters.  

Lorentz symmetry claims that there is no preferential inertia frame. 
However,  our actual universe is filled with the CMB and cosmic matter 
and  we can clearly identify the preferential frame, which we have called 
the C-frame. Since this C-frame is approximately identical with CMB 
isotropic frame, the C-frame  is supposed to be selected during the 
reheating phase at the Inflation, that is, in association with some SSB 
of  vacuum state in quantum field theory. Some physical parameters of 
particles in "our universe" is supposed to have been built through this 
dynamical process  of  the SSB.  

Our speculation in this report is extend this SSB paradigm into  
the creation of spacetime of "our universe". If the spacetime of 
"our universe" is a product of the  various dynamical processes, it 
would be very natural to expect a breakdown of original symmetry of 
spacetime. That is a spontaneous breaking of exact Lorentz symmetry. 
According to some Toy models of the SSB for Lorentz symmetry, we could 
suppose that each particle gets their different limiting velocity: the 
maximum velocity at infinity energy for massive particle.  We will 
discuss such Toy model later in this report, where the SSB is implemented 
by an "invisible un-removable"  field.

Lorentz symmetry, however, has been built in all fundamental concepts of  
modern physics, such as  Dirac field, spin, renormalization group of quantum 
field theory, and so on. Therefore, the violation of this symmetry can not 
be introduced so easily. One of the claim of the relativity principle is 
the equivalence of all inertia frame. However this equivalence has not been 
directly proved so much. Only the accelerator experiments has proved this 
equivalence  up to some Lorentz factor of $\gamma_{\rm acce} \sim 10^5$. 

In this respect, the GZK cut-off~\cite{bgz} has an unique status for the 
experimental verification of this equivalence. In 1972, I discussed this 
point and wrote a paper  with the title "Ultra-high Energy cosmic rays, Hot 
universe and Absolute reference frame".~\cite{b72} By this GZK cut-off, we 
will be possible to check the validity limit of the relativity principle  
up to $\gamma_{\rm GZK} \sim 10^{11}$. Following to the above Toy model, 
this verification can be regarded as a check of the universality of the 
limiting velocity. And if there  were not the GZK cut-off, that may imply 
a finding of a un-removable hidden external field of vector or tensor type 
external field in addition of "scalar" type  Higgs field.

\section{Comoving Frame in the Expanding Universe and Relativity Principle}

In our expanding universe, we can easily identify  preferential inertia 
frames: (1)rest frame of baryon matter, (2)rest frame of astronomical 
objects, (3)frame in which CMB is isotropic, (4) frame in which the Hubble 
flow is observed isotropic. Furthermore, these four frames are approximately 
identical within a relative velocity difference of several hundreds km/sec. 
These inertia frames have a concrete physical effect in a process of the 
structure formation in the expanding universe.~\cite{{b98},{b01},{b03}} 

According to theoretical view, these cosmological frames are considered to 
have the same physical origin; spontaneous selection of the inertia frame at 
the SSB of the quantum vacuum. However, even in the vacuum universe without 
material substance, the creation of the expanding universe itself is  the 
browken state of  Lorentz invariance. That is a formation of comoving frame  
perpendicular to the time direction. We call this cosmological and comoving 
frame as C-frame.

In spite of a lucid existence of the C-frame, however, the Lorentz symmetry  
is supposed to hold in any local physical phenomena. The relativity principle 
does not respect this lucid existence. That is the spirit of the relativity 
since Galileo. In the derivation of GZK cut-off, the relativity principle 
is used as usual but its situation is very special because the Lorentz 
factor relative to the C-frame is as large as $\gamma \sim 10^{11}$, which 
is far beyond the Lorentz factor in the particle accelerators of about 
$\gamma \sim 10^5$.

Here we should not confuse the two meanings of "high energy".
One is an invariant energy(or center of mass energy) defined such as ,
$$p^\mu p_\mu=E^2-P^2=Q^2$$, where $p^\mu$ is total four momentum of the 
system. 
Another one is  energy relative to a specific reference frame and it will 
be defined in the following manner as
$$ N^\mu p_\mu=1\cdot E-0\cdot P=E$$
, where $N^\mu$ is the  four vector which specifies this particular reference 
frame. For the C-frame in the expanding universe, the component is given as 
$N^\mu(1,0,0,0)$ . The Relativity principle claims that the cross section of 
collision, $\sigma$, does depend solely on $Q$ but does  not depend on 
$N^\mu p_\mu$ , such as $\sigma(Q)$ but not as $\sigma(Q, N^\mu p_\mu)$.  
In my early paper~\cite{b72}, the cut-off function in the momentum space 
was assumed to depend on $N^\mu p_\mu$ and the cross section involved to 
the GZK cut-off could be modified; the cut-off in the momentum space does 
affect the final state integral.

The GZK cut-off, $Q$ is $\sim 10^{8.5}$eV, which is rather low energy in  
high-energy physics, but, $N^\mu p_\mu \sim 10^{20}$eV is extraordinarily 
large even in high-energy physics. The uniqueness of the GZK cut-off lies 
on the largeness of $N^\mu p_\mu$. Therefore, we should not confuse with  
the so-called energy frontier in the high-energy physics, e.g., Energy 
frontier for supersymmetry, GUT, Planck scale, etc..Those are talking about 
large $Q$ but not on the largeness of $N^\mu p_\mu$.  

\section{A Toy Model for Violation of Lorentz Symmetry}

We discuss a Toy model which introduces the differentiation of the universal 
limiting velocity  by the SSB.~\cite{b03}  Consider the following Lagrangian 
for a Dirac particle A,

$$ L_A={i \over 2}\bar{\psi}\gamma_\mu \partial^\mu \psi-\alpha_A \phi 
\bar{\psi}\psi+{i \over 2}g_A F_{\mu \nu}\bar{\psi}\gamma^\mu\partial^\nu 
\psi,$$

where $\psi$ is the Dirac field of A-particle, $\phi$ is Higgs scalar field 
with coupling coefficient $\alpha_A$ and $F_{\mu \nu}$ is a tensor field with 
coupling coefficient $g_A$. The first term in the right hand side is kinetic 
term and the second one is the Yukawa coupling term which creates mass by 
Higgs mechanism.  In this Lagrangian, the dynamical parts of $\phi$ and 
$F^{\mu \nu}$  has been omitted and $\phi$ and  $F^{\mu \nu}$ are both taken 
as an external field. These external fields are "un-removable"  from our 
universe. Non-zero value of $<\phi>$ gives the mass, $m_A=\alpha_A <\phi>$, 
to this Dirac particle.

Next we assume that some component of the tensor field has got some non-zero 
value as followings,

$$<F^{00}>=B\neq 0~~ {\rm and} <F^{\mu\nu}>=0~~ 
{\rm for~~other~~components}.$$ 
$B$ is supposed to be nearly constant in conventional physical scale of 
spacetime, but it can be slowly changing with cosmological spacetime scale. 
This $B$ is "un-removable" field. 
Then the dispersion relation for plain wave is given as~\cite{bbt}

$$p^\mu p_\mu -m_A^2c^2=-2g_AB(E/c)^2$$, 
where only the first order terms of B has been retained and the higher term 
of B has been neglected.

This relation is rewritten by denoting the three momentum as $p$ as
$$(1+g_AB)(E/c)^2=p^2+m_A^2c^2,$$
where $c$ is the universal constant introduced at the definition of the 
spacetime length by space length and time length.

Renormalizing the velocity and mass as followings
$$ c_A^2={ c^2 \over 1+g_A B}~~ {\rm and}~~m_{AB}^2=(1+g_AB)m_A^2 ,$$
the conventional energy-momentum relation is resumed~\cite{b03}

$$ E^2=p^2 c_A^2 + m_{AB}^2c_A^4.$$
but now $c_A$ is depending on particle species through $g_A$, that is, the 
limiting velocity, velocity in the limit of $E \rightarrow \infty$, is 
depending on the particle species.

Here we remark some difference between the Higgs scalar $\phi$ and the 
tensor external field $F^{\mu \nu}$. Since the value of tensor components 
is dependent on the choice of the inertia frame, we have adopted the C-frame 
as the preferential frame and the above energy-momentum relation holds only 
in the C-frame. This is different from the case of a scalar field, where the 
value of external field is independent from the inertia frame. 

If we modified the Lorentz transformation with psudo-Lorentz factor
$$ \gamma_A={1 \over \sqrt{1-\left({ v \over c_A}\right)^2}}~~
{\rm instead~~of~~} \gamma={1 \over \sqrt{1-\left({ v \over c}\right)^2}},$$
the above relation keeps its form. However the Lorentz invariance apparently 
breaks down if we consider a system consisting of pariticles of different 
species. That is, Violation of Lorentz symmetry has been introduced through 
the differentiation of the limiting velocity for each species of particles.

The perturbative super string theory has suggested an existence of various 
hidden fields such as the above tensor field. 
If we assume a vector field $A_\mu$ in stead of $F_{\mu \nu}$ as the external 
field, the Lagrangian is written,~\cite{bkl}
$$ L_A={i \over 2}\bar{\psi}\gamma_\mu \partial^\mu \psi-m_A \bar{\psi}
\psi-f_AV_\mu \bar{\psi}\gamma^\mu  \psi.$$
, where the Higgs term is now rewritten by the mass term.
Here we assume 
$$ <V_0>=V\neq 0 ~~{\rm and}~~<V_\mu=0>~~{\rm for~~all~~other~~ components}$$
This $V$ is also "un-removable" field. The the dispersion relation becomes 
like
$$E^2-p^2c^2-m_A^2c^4=-2f_AVE.$$
If we define as
$$c_A(E)={ c \over 1+{f_AV \over E}},~~ m_{AV}^2=(1+f_AV/E)^2 
[m_A^2+(f_A V)^2/c^4],$$
the above dispersion relation resume  a pseudo-conventional form like
$$E^2=p^2c_A(E)^2+m_{AV}^2c_A(E)^4.$$
$c_A(E)$ has anomaly in the limit of $E \rightarrow 0$ but this limit would 
need a quantum mechanical correction. The violation of Lorentz invariance 
would dominate in the vector case similar to the scalar or Higgs case. Then 
the tensor case is necessary as the toy model which exhibits the violation 
of Lorentz invariance in the limit of large $\gamma$

\section{"Miss Conduct" of Lorentz Transformation}

Here we mention the two types of Lorentz transformation, those are often 
comfused each other. One is the boost particle-transformation and the another 
one is the Lorentz transformation.~\cite{bck} The Lorentz transformation is 
just a change of reference frame for the description of the same phenomena, 
whic is sometime called "passive" transformation. In contrast to this, the 
boost particle-transformation  is "active" transformation, where particle's 
energy-momentum are changed actually. Relativity principle claims that the 
boosted state and the original state seen from the transformed reference 
frames are identical. For the system of particles, this classification has 
no particular useful meaning.

However some complication comes in, when  the system consists of  particles 
and external given field. In the Lorentz transformation, both the  particle's 
energy-momentum and the components of the external field are both 
transformed. Therefore the relative relation between particle and external 
field does not changed. In the boost particle-transformation, however, 
particle's energy-momentum are transformed but the field configuration is 
kept unchanged. Therefore particle states  relative to the field are 
different. The actively boosted state of particle is not identical with the 
passively Lorentz transformed state having the same particle state but 
different field configuration. Thus we call this situation as an "apparent" 
violation of Lorentz invariance. We can even say that  the boost 
particle-transformation is in fact a misconduct of the Lorentz transformation.

What we have done in the previous section is something like this misconduct. 
In the actual universe, the external fields like $F^{\mu \nu}$ are totally 
unknown to us upto now and "misconduct" of application of the Lorentz 
transformation could happen. Conversely we also say that the apparent 
violation implies a finding of the hidden  external fields and the genuine 
Lorentz symmetry for the total system(particle and external field) could 
recover.

\section{Different Limiting Velocities and GZK cut-off}

Without touching on any  origin of different limiting velocities, we can 
rise a question how much degree the universality of limiting velocity has 
been checked by direct experiment. The assumption of non-equality of the 
limiting velocity of a charged particle and light velocity  is equivalent 
to the  introduction of the Lorentz non-invariant term of the electromagnetic 
field into the Lagrangian.~\cite{bgd} In general, this is  true for any  
non-universal assumptions of the limiting velocity.~\cite{bc9}

Coleman and Glashow also discussed this assumption, firstly in order to 
explain the neutrino oscillation.~\cite{bc7} They also pointed out that the 
high-energy phenomena might disclose  an apparent degeneracy of limiting 
velocity and reveal  a splitting into a fine structure of the limiting 
velocity. They called various limiting velocity as eigen state of velocity 
for each particle. They have shown also that this modification does not hurt 
the standard theory of interaction based on the gauge field theory.~\cite{bc9}

If the limiting velocity is dependent on the particle species A like  $c_A$, 
the GZK cut-off discussion is  modified very much. By the head-on collision 
between the cosmic-ray proton and the CMB photon, $\Delta$ particle is 
produced if the following condition is satisfied.~\cite{bbt}
$$(E_p+E_\gamma)^2-(p_p+p_\gamma)^2c_\Delta^2 >m_\Delta^2 c_\Delta^4,$$
while the proton obeys to $E_p^2=p_p^2c_p^2+m_p^2c_p^4$. In the situation 
of $E_p \gg m_p c_p^2$ and $\vert c_\Delta-c_p \vert\ll c_p$, the  condition 
becomes as followings
$$-{c_\Delta-c_p  \over c_p}E_p^2+2E_pE_\gamma >{m_\pi^2c^4 \over 2}$$
In the conventional case, $c_\Delta-c_p =0$ and the threshold energy is 
obtained $E_p>m_\pi^2 c^4/4 E_\gamma$. 

If $(c_\Delta-c_p)\neq 0$, the above equation gives a quite different result; 
the cut-off  disappears for $(c_\Delta-c_p)>0$ and the cut-off energy  
decreases compared with the GZK cut-off for $(c_\Delta - c_p)<0$. For 
example, the above equation does not have solution if
$${c_\Delta-c_p \over c_p} >2\left({E_\gamma \over m_\pi c^2}\right)^2
\sim 10^{-22},$$
the cut-off does not exist.

On the other hand, for $(c_\Delta-c_p)<0$, the cut-off energy is modified as
$$E_{GZK}\left[1- {\vert c_\Delta-c_p\vert \over 2c_p}\left({m_\pi c^2 
\over E_\gamma}\right)^2\right]~~{\rm for}~~{\vert c_\Delta-c_p\vert \over 
2c_p}\left({m_\pi c^2 \over E_\gamma}\right)^2 < 1$$
and 
$$ \sqrt{c_p \over 2\vert c_\Delta-c_p \vert} m_\pi c^2~~{\rm for}~~
{\vert c_\Delta-c_p \vert \over c_p}m_\pi^2 c^4 \gg E_\gamma^2$$

\end{document}